\documentclass[aps,pre,twocolumn,groupedaddress,showpacs,amsmath,amssymb,fleqn,floatfix]{revtex4}
\usepackage{graphicx}

\begin{document}

\title{Poisson Nernst-Planck Model of Ion Current Rectification\\ through a Nanofluidic Diode}

\author{D. Constantin}
\email[]{dragos@uci.edu}

\author{Z. S. Siwy}
\email[]{zsiwy@uci.edu}

\affiliation{Department of Physics and Astronomy, University of California, Irvine, CA 92697}

\date{\today}

\begin{abstract}
We have investigated ion current rectification properties of a recently prepared bipolar nanofluidic diode. This 
device is based on a single conically shaped nanopore in a polymer film whose pore walls contain a sharp boundary 
between positively and negatively charged regions. A semi-quantitative model that employs Poisson and Nernst-Plank 
equations predicts current-voltage curves as well as ionic concentrations and electric potential distributions in 
this system. We show that under certain conditions the rectification degree, defined as a ratio of currents recorded 
at the same voltage but opposite polarities, can reach values of over a $1000$ at a voltage range $<-2\,V, +2\,V>$. 
The role of thickness and position of the transition zone on the ion current rectification is discussed as well. We 
also show that rectification degree scales with the applied voltage.
\end{abstract}


\maketitle


\section{\label{intro}Introduction}


Nanopores in polymer films and silicon materials have attracted a lot of scientific interest due to their application 
as biomimetic systems for models of biological channels \cite{Lev:1993rs, Siwy:2003ps, Chen:2004al}, and as 
biosensors \cite{Li:2001ib, Dekker:2007ss, Harrell:2006rp, Mara:2004aa}. Nanopores are also used as a basis for 
building ionic devices for controlling flow of ions and charged molecules in a solution \cite{Gijs:2007}. In these 
systems electrostatic and hydrophobic interactions between the ions and molecules that pass through the pore and the 
pore walls are used to amplify or stop the transport. Our group has recently prepared a nanofluidic diode 
\cite{Vlassiouk:2007} which rectifies ion current in a similar way as a bipolar semiconductor diode rectifies 
electron current. This diode is based on a single conically shaped nanopore in a polymer film with openings of 
$2\div6\,nm$ and $1\,\mu m$, respectively. The surface charge of the pore is patterned so that two regions of the 
pore with positive and negative surface charges create a sharp barrier called the transition zone. This nanofluidic 
diode is bipolar in character since both positively and negatively charged ions contribute to the measured current. 
Another type of nanofluidic diode was prepared in the group of A.\ Majumdar \cite{Karnik:2007ri} with a sharp barrier 
between a positively charged zone and a neutral part of the pore. Presence of only one type of surface charges causes 
the latter device to be unipolar.

In this article we provide a semi-quantitative description of ion current rectification of the bipolar diode based on 
single conical nanopores. We use continuum mean-field theory, and more precisely Poisson and Nernst-Planck (PNP) 
equations, to define our model \cite{Schuss:2001dp, Nadler:2004id, Chen:1997pt, Liu:2007ap} that we apply to describe 
current-voltage characteristics as well as profiles of ionic concentrations and electric potential in the system. Our 
model is based on a similar approach as described by Cervera et al. in \cite{Cervera:2005, Cervera:2006}. We would 
like to mention that the model of nanofluidic diode based on a rectangular channel with limiting length of tens of 
nanometers  presented in \cite{Daiguji:2005nd} combines PNP with Navier-Stokes equations. Nanopores considered here 
are significantly narrower with the smaller diameter around $5\,nm$. Therefore, we consider here only PNP equations, 
because in nanopores of diameters less than $10\,nm$, electrophoresis is known to be the most dominant process 
\cite{Daiguji:2004it, Karnik:2007ri}.

It is important to mention that continuum description of ion transport by the PNP equations has successfully rendered 
main properties of many biological channels, e.g.\ see \cite{Nonner:1998} for the calcium channels treatment. When 
compared to more precise but computational expensive techniques, e.g.\ Brownian dynamics \cite{Corry:2000}, the 
continuum approach has a clear advantage in terms of computing cycles. As discussed in \cite{Corry:2000} this 
description produces results which correctly render the physical and chemical phenomena occurring inside a 
cylindrical nanopore whose diameter is bigger than $2$ Debye lengths of the considered ionic species. Hence it is 
plausible to obtain valid results for the conical nanopores as long as their smallest diameter is bigger than $2$ 
Debye lengths in the examined system.

In this study, we thoroughly discuss how ion current rectification in a nanofluidic diode depends on the position of 
the transition zone between the two regions of pore walls with opposite surface charges. Influence of the width of 
the transition zone on rectifying properties of the device is studied in detail as well. We identify the range of 
values for the position of the transition zone and its width that lead to a significant enhancement of the current 
rectification degree. This aspect is of crucial importance for further improvement of these new devices. The paper is 
organized as follows. In section \ref{s:methods} we explain the experimental system and present the theoretical basis 
of our model. In section \ref{s:results} we present the results of the analysis and provide directions to future 
improvements of the design of nanofluidic diodes.


\section{\label{s:methods}Methods}


In this section we present the experimental results and the theoretical background which provide the basis of our 
simulations. In the first subsection we briefly explain the experimental facts which inspired the current project. 
The following subsections are devoted to the theoretical background.

\subsection{\label{experimental}Experimental}

Single nanopores in polymer (polyethylene terephthalate, in short PET) film were prepared by the track-etching 
technique as described in \cite{Apel:2001dl, Siwy:2006ic}. Briefly, the method consists of irradiation of polymer 
films with single swift heavy ions \cite{Spohrpatent} (Gesellschaft fuer Schwerionenforschung, Darmstadt, Germany), 
and subsequent asymmetric etching of the irradiated foils at $9 M$ $NaOH$ \cite{Apel:2001dl}. This procedure leads to 
preparation of membranes that contain only one pore with the small opening, called tip, as narrow as several 
nanometers in diameter, and the big opening, called base, with micrometer size diameter.
\begin{figure}[ht]
\includegraphics[width=0.48\textwidth]{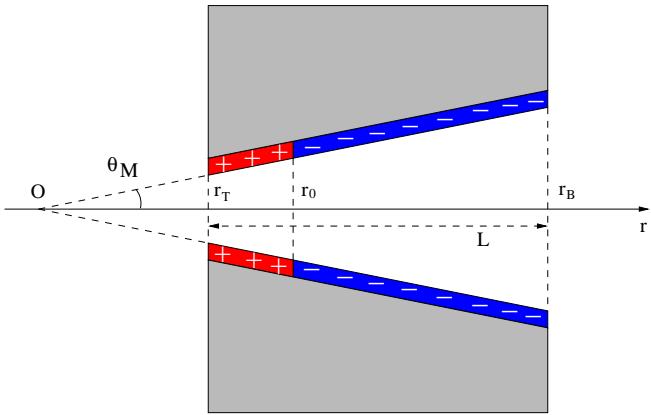}%
\caption{\label{fig:pore-geometry}Pore geometry with schematic representation of surface charge distribution creating 
a bipolar nanofluidic diode.}
\end{figure}
As the result of the chemical etching, there are carboxyl groups formed on the surface of pore walls with density of 
about one per $nm^{2}$, which determines the surface charge to $-0.16\,C/m^2$ \cite{Wolf}. In order to modify the 
surface of PET nanopores we used the method of coupling carboxyl and amine groups with 
1-Ethyl-3-[3-dimethylaminopropyl]carbodiimide hydrochloride (EDC) \cite{Grabarek:1990zl}. The reaction between the 
carboxyl groups and diamines changes the surface charge from negative to positive. Targeted modification of the tip 
was obtained by introducing the reagent mixture only on the side of the membrane with the small opening 
\cite{Vlassiouk:2007}. The other side of the membrane was in contact with a buffer solution. A conical shape of the 
pore makes the distribution of the diamines and EDC in the pore extremely non-homogeneous. There is a high 
concentration of the reagents at the tip of the pore that assures the chemical modification reaction of the tip to 
occur. The concentration of diamines and EDC at the remaining part of the pore is very low, therefore this part 
remains negatively charged. As the result, we obtain a surface charge distribution as schematically shown in 
Fig.~\ref{fig:pore-geometry}. An example of the current-voltage characteristic through such a device recorded at 
$0.1\,M$ $KCl$ and $pH\,5.5$ is presented in Fig.~\ref{fig:iv-exp}. The rectification degree of the system is defined 
as the ratio of currents recorded for positive and negative voltages, respectively
\begin{equation}
\label{rectif-deg}
f(V)=\frac{|I(V)|}{|I(-V)|}\,.
\end{equation}
The rectification degree of this system (Fig.~\ref{fig:iv-exp}) equals $217$ at $5\,V$ \cite{Vlassiouk:2007}. We 
would like to mention that even devices with uniform surface charge distribution rectify the current but their 
rectification degree is one order of magnitude smaller than the one obtained for the systems described above 
\cite{Apel:2001dl, Siwy:2003ps}.
\begin{figure}[ht]
\includegraphics[width=0.48\textwidth]{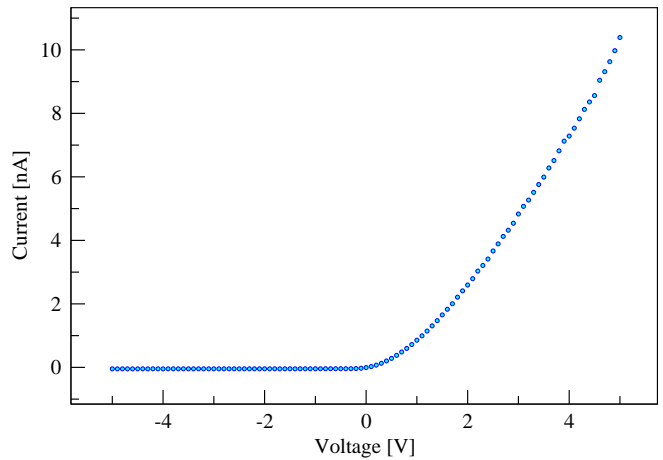}%
\caption{\label{fig:iv-exp}Experimental data of ion current recorded at $0.1\,M$ $KCl$, $pH\,5.5$ through a single 
conical nanopore with surface charge distribution as shown in Fig \ref{fig:pore-geometry} \cite{Vlassiouk:2007}. The 
pore diameters were $5\,nm$ and $1000\,nm$, respectively}
\end{figure}
It is also important to mention that the single pore diode is a nano-realization of bipolar membranes consisting of 
cationic and anionic membranes \cite{Coster:1965qa, Bassignana:1983it, Mafe:1997ec}.

\subsection{\label{ss:m}Modeling}

Before we start to introduce the theoretical base of the model we would like to specify the notation. We have 
assigned the subscripts ``$T$" and ``$B$" to designate the various physical quantities computed at the tip and at the 
base sides of the pore, respectively. In this sense $r_T$ is the location of the pore tip and $r_B$ is the location 
of the pore base. The thickness of the membrane is denoted with $L$. The cone opening angle is $\theta_M$. The 
membrane is immersed in an electrolyte containing potassium and chloride ions, and in addition we apply an external 
electric field with the help of two electrodes positioned far away from the membrane on each side of the pore. Due to 
$G\Omega$ resistance of single nanopores, the voltage drop will occur only inside the pore, not in the electrolyte 
bulk. By convention, we will keep the electrode on the pore tip side grounded such that the sign of the electric 
potential difference will be controlled by the electrode on the base side of the pore. The existence of the external 
field causes the ions to migrate towards the electrodes, which we record as ion current. We would like to mention 
that for small pores \cite{Siwy:2003ps} and more complex electrolytes in which chemical reactions can occur, one 
observes large fluctuations of ion current in time \cite{Siwy:2006ni, Siwy:2006ci}. However our model in its present 
form does not include chemical reactions between components of the electrolyte, and therefore the system will reach a 
stationary regime, i.e., the ionic fluxes are constant in time. In nanofluidic diodes, constant currents at constant 
voltages are indeed observed experimentally. Our goal is to compute the steady state current for a fixed set of 
physical and chemical parameters. Hence we express the stationarity of the ionic fluxes using Nernst-Plank condition. 
To be more specific we assume that for each ionic species, indexed by ``$i$'', the molar flux defined as
\begin{equation}
\label{molar-flux}
\vec{J}_i(\vec{r}) = -D_i \big[~\nabla c_i(\vec{r}) + z_i c_i(\vec{r}) ~\nabla \phi(\vec{r})~\big] \,,
\end{equation}
is conserved, i.e., it obeys the Nernst-Plank condition
\begin{equation}
\label{NP}
\nabla \vec{J}_i(\vec{r}) = 0 \,,
\end{equation}
where $D_i$ is the diffusion coefficient of an ion $i$, $c_i$ is the molar concentration and $z_i$ is the charge 
number of the ion $i$. We have denoted with $\phi$ the electric potential in $RT/F$ units, which satisfies the 
Poisson equation
\begin{equation}
\label{P}
\triangle \phi(\vec{r}) = -\frac{F^2}{\varepsilon RT} \sum_i z_i c_i(\vec{r}) \,,
\end{equation}
where $\varepsilon$ is the electric permittivity of the medium and the summation is carried over all the ions present 
in the solution. The parameter $F$ is the Faraday constant, $R$ is the gas constant and $T=293\,K$ is the absolute 
temperature. The PNP equations \eqref{NP} and \eqref{P} provide a complete set of equations which together with 
appropriate boundary conditions allow us to determine the concentration of each ionic species and the electric 
potential inside the pore. Consequently the total ion current density along an arbitrary direction $\hat{n}$ can be 
computed with the help of \eqref{molar-flux}
\begin{equation}
\label{current-density}
j_n(\vec{r}) = F \sum_i z_i \vec{J}_i(\vec{r}) \cdot \hat{n}\,,
\end{equation}
where the summation is performed again over all considered ions.

As the problem has azimuthal symmetry with respect to the pore axis, it gets reduced to a set of two-dimensional 
second order partial differential equations. To simplify the problem even more we perform an analytical integration 
with respect to the polar angle and we obtain a set of second order ordinary differential equations which is finally 
solved numerically. We would like to mention that after integration one has to neglect one term in order to obtain a 
well defined set of ordinary differential equations \cite{Cervera:2006}. Our comparison tests between the full two 
dimensional problem and the approximated one dimensional problem show insignificant differences in both the 
quantitative and the qualitative results. Hence the solution of the problem will provide averaged quantities in the 
sense that the polar angle dependence is lost during the integration. However our main interest is to compute the ion 
current, which involves averaging along the polar coordinate, therefore we do not loose any information. To be more 
specific, the differential equations which result from the above described procedure have the following structure
\begin{equation}
\label{NP1}
\frac{d}{dr}\Big[ r^2 \Big( \,\frac{dc_i(r)}{dr} + z_i c_i(r)\, \frac{d\phi(r)}{dr} \,\Big) \Big] = 0 \quad\quad 
\forall i\,,
\end{equation}
and
\begin{equation}
\label{P1}
\frac{1}{r^2}\frac{d}{dr}\Big(r^2 \frac{d\phi(r)}{dr}\Big)=-\,\frac{F^2}{\varepsilon RT}\Big[\sum_i z_i c_i(r) + 
X(r)\Big] \,,
\end{equation}
where the quantities $c_i(r)$ and $\phi(r)$ in \eqref{NP1} and \eqref{P1} come from integration, i.e.,
\begin{subequations}
\label{mediated}
\begin{align}
\phi(r) &=\frac{1}{1-\cos\theta_M}\int_0^{\theta_M} d\theta \,\sin\theta \,\phi(r,\theta) \,,\label{phi-med}\\
c_i(r) &=\frac{1}{1-\cos\theta_M}\int_0^{\theta_M} d\theta \,\sin\theta \,c_i(r,\theta) \,,\label{c-med}
\end{align}
\end{subequations}
and the function $X(r)$ is given by
\begin{equation}
X(r) = \frac{\sin\theta_M \sigma(r)}{Fr(1-\cos\theta_M)} \,,
\end{equation}
where $\sigma(r)$ represents the surface charge density. For simplicity we have kept the same notation for the 
averaged quantities $c_i(r)$ and $\phi(r)$ and we will use the functional dependence on the polar angle to 
distinguish between them and the non-averaged quantities $c_i(r,\theta)$ and $\phi(r,\theta)$.

\subsection{\label{ss:bc}Boundary Conditions}

To completely define the problem we have to specify the boundary conditions. These conditions are derived from the so 
called Donnan equilibrium conditions which express the thermodynamic equilibrium between the membrane with excess 
surface charge and the ionic concentrations in the bulk. One can derive these conditions as in \cite{Gillespie:2001} 
and obtain the following relations for the concentrations at the pore border
\begin{equation}
\label{conc}
c_i(r_j) =c_{0i}\,\exp[-z_i\,\phi_{D}(r_j)] \,,
\end{equation}
where $c_{0i}$ represents the bulk concentration of ion $i$ and $j=T,B$ indexes the tip and the base sides of the 
pore, respectively. $\phi_{D}$ is the Donnan potential defined as the potential difference created across an ion 
exchange membrane
\begin{equation}
\label{donnan}
\phi_{D}(r_j) = \phi(r_j) - \phi_j\,,
\end{equation}
where $\phi_j$ is the electric potential at the corresponding electrode. The electro-neutrality condition, which has 
the form
\begin{equation}
\label{enr}
\sum_i z_i c_i(r_j) +X(r_j) = 0 \,,
\end{equation}
provides an equation for $\exp[\phi_{D}(r_j)]$. Hence, the boundary conditions can be derived from \eqref{conc} and 
\eqref{enr} as
\begin{equation}
\label{c-bc}
c_i(r_j) =\frac{1}{2}\Big( -z_i X(r_j) + \sqrt{X(r_j)^2+4 c_0^2} \,\Big)\,,
\end{equation}
and
\begin{equation}
\label{p-bc}
\phi(r_j) = \phi_j - \frac{1}{z_i} \ln{\frac{c_i(r_j)}{c_0}} \,.
\end{equation}
The PNP equations \eqref{NP1} and \eqref{P1} together with the boundary conditions \eqref{c-bc} and \eqref{p-bc} form 
the base of our model. We deal with two ionic species in our system, i.e., $K^+$ and $Cl^-$, therefore we have to 
solve a set of three second order ordinary differential equations.

\subsection{\label{ss:scd}Surface Charge Distribution}

The charge distribution is formed by two distinct regions with positive and negative surface charges respectively, 
and a transition zone between them. These regions are characterized by the asymptotic values of the surface charge 
density, i.e., the value of $\sigma$ far away from the transition zone. We employ some simplifying assumptions in 
order to characterize the surface charge distribution in the transition zone. Firstly, we consider that the 
magnitudes of the asymptotic values of the positive and negative surface charge densities are the same and equal to 
$e_0/nm^2$, where $e_0$ is the elementary charge. Additionally, we assume that the transition between the positive 
and negative charges on pore walls is made in a symmetric, continuous, and monotonic way.
\begin{figure}[ht]
\includegraphics[width=0.48\textwidth]{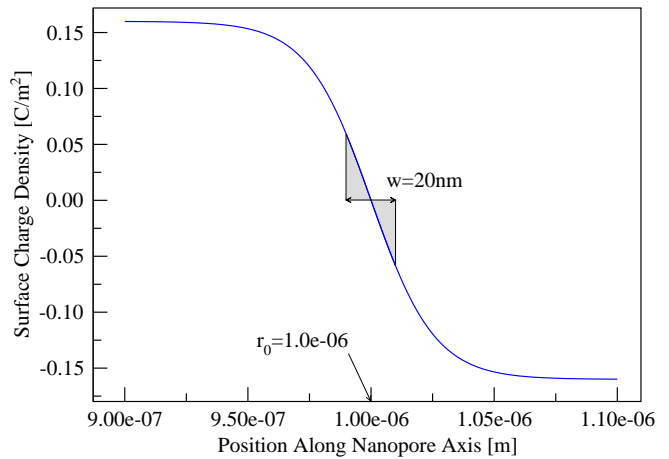}%
\caption{\label{fig:surf-charge}Surface charge distribution that has been used in calculations. The width of the 
transition zone $w$ and position of the zero charge point $r_0$ are indicated in the figure.}
\end{figure}
Hence we will consider the surface charge density in the pore to be described by the logistic type function (see 
Fig.~\ref{fig:surf-charge}):
\begin{equation}
\label{surface-charge}
\sigma(r)=\sigma_0\,\bigg\{ -1 + \dfrac{2}{1+\exp[-2k(r-r_0)]}\bigg\} \,,
\end{equation}
where $\sigma_0=-e_0/nm^2$ is the minimum value of the surface charge density, $r_0$ is the position along the pore 
axis where $\sigma(r)$ equals zero, and $k$ measures the slope of $\sigma$ in the transition zone. Another equivalent 
way to characterize the charge distribution in the transition zone is to specify the transition zone width $w$ (see 
Fig.~\ref{fig:surf-charge}) which is defined as the part of the pore where the magnitude of the charge is bellow 
$|\sigma_0|/e$, i.e.,
\begin{equation}
\label{width}
w=\frac{1}{2k}\ln\frac{m_2}{m_1} \,,
\end{equation}
where $m_{1,2}=\tanh^{\pm1}(1/2)$ are constants. In this problem there are many parameters which can be modified. 
However our main interest was to see how the properties of the nanofluidic diode modify when the surface charge 
distribution parameters change. Therefore, we have chosen to keep all the parameters constant, the only exception 
being $r_0$ and $w$, the parameters, which we believe can be controlled experimentally.


\section{\label{s:results}Results}


As we have discussed in the previous section we have to solve a boundary value problem (BVP) defined by a set of 
three second order ordinary differential equations. The solution will provide the potassium and chloride
concentrations and the electric potential as functions of the coordinate along the pore axis. We have used the BVP 
solver described in \cite{Shampine:2006} to compute the solution with an error less than $10^{-6}$. The solution of 
the computation is used to evaluate the ion current which is given by
\begin{equation}
I = 2\pi (1-\cos\theta_m)\,r^2 j_r(r) \label{current}\,,
\end{equation}
where $j_r$ is the radial component of the ion current density given by \eqref{current-density}. As we expect the 
current is not dependent on the radial coordinate and it has a constant value along the pore axis.

\subsection{\label{ss:cv}Current-voltage curves and ionic concentration profiles}

\begin{figure*}[ht]
\includegraphics[width=\textwidth]{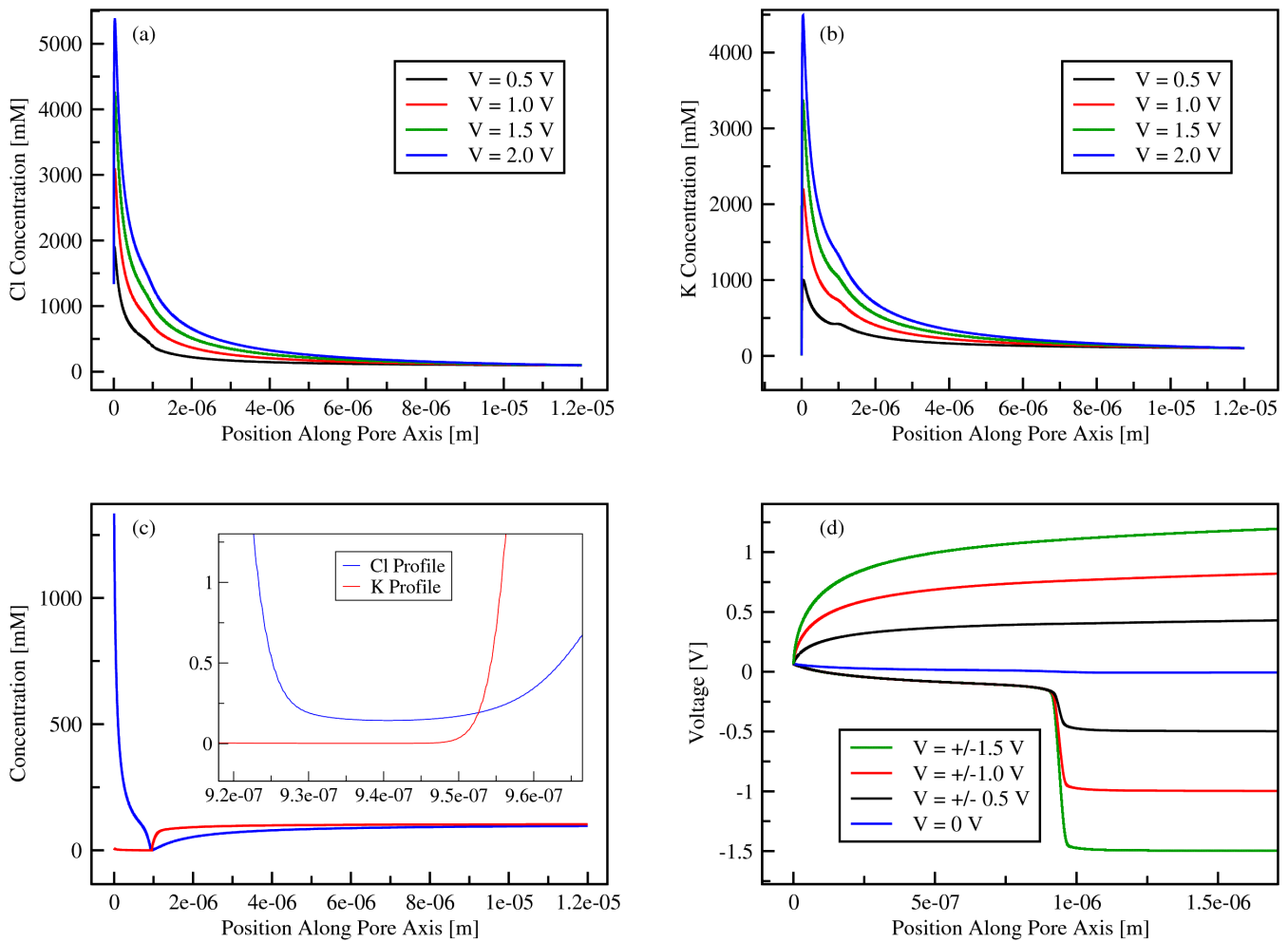}%
\caption{\label{fig:profiles}Profiles of ionic concentrations and electric potential in a nanofluidic diode based on 
a single conical nanopore for $r_0=1\,\mu m$, $w=100\,nm$, and various voltages as indicated. (a) profiles of $Cl^-$ 
concentration in a nanopore for forward bias (positive voltages); (b) $K^+$ concentration profiles for forward bias; 
(c) profiles of $Cl^-$ and $K^+$ concentrations for reverse bias at $-1\,V$; (d) electric potential profiles in the 
nanopore for forward and reverse bias.}
\end{figure*}

\begin{figure}[ht]
\includegraphics[width=0.48\textwidth]{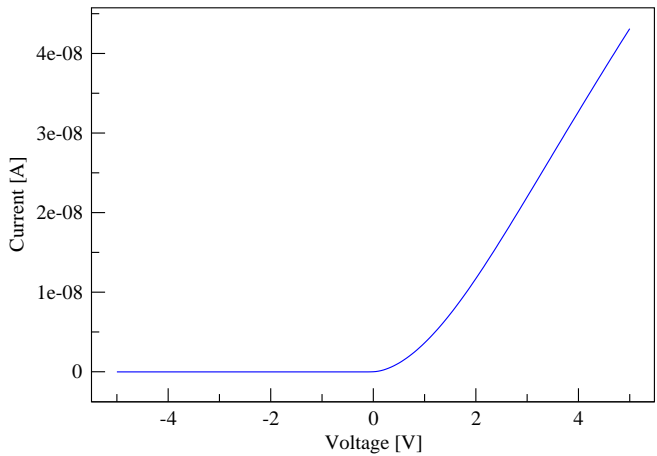}%
\caption{\label{fig:iv-curve}Current-voltage curve for a nanofluidic diode calculated with Eq.~\eqref{current} for 
$r_0=100\,nm$ and $w=100\,nm$. The surface charge was assumed as shown in Fig.~\ref{fig:surf-charge}. The pore 
openings were $5\,nm$ and $1000\,nm$, respectively.}
\end{figure}

The system that we have considered is a single conical nanopore with pore diameters $5\,nm$ and $1000\,nm$, 
respectively, at symmetric electrolyte conditions of $100\,mM$ $KCl$ on both sides of the membrane. The thickness of 
the membrane is $12\,\mu m$.  Fig.~\ref{fig:profiles}(a-c) show the numerical solutions of the corresponding 
distributions of potassium and chloride ions in forward and reverse bias, respectively. Fig.~\ref{fig:profiles}(d) 
presents the distribution of the electric potential in the nanopore for various applied voltages. The increase of 
ionic concentrations in the forward bias is remarkable. Concentrations of both ions, potassium and chloride, have 
increased by an order of magnitude, compared to the bulk concentration of $100\,mM$. This result also confirms 
bipolar character of our device, since $K^+$ and $Cl^-$ contribute almost equally to the measured ion current. Ionic 
concentrations at the reverse bias are dramatically different. We see significantly lower concentrations of both ions 
in a region close to the transition zone which indicates the formation of the depletion zone. The inset in 
Fig.~\ref{fig:profiles}(c) shows a magnified view of the depletion zone where the concentration of both ions drops 
practically to zero. This region is located slightly off the center of the transition zone because of the conical 
geometry of the nanopore. Fig.~\ref{fig:profiles}(d) shows profiles of electric potential in the pore for various 
voltages applied in the forward and reverse directions. Similar to the results for diodes based on rectangular 
channels and bipolar membranes \cite{Karnik:2007ri, Daiguji:2004it, Bassignana:1983it, Mafe:1997ec}, in our bipolar 
diode based on single conical nanopores, the whole voltage drop occurs in the depletion zone. This sudden drop of the 
electric potential causes the appearance of a huge electric field located in the depletion zone, which will 
effectively block the ionic currents in the reverse bias.

Using the numerical solution and Eq.~\eqref{current} we compute the ion current. Fig.~\ref{fig:iv-curve} shows a 
typical current-voltage curve calculated for this nanopore with parameters $r_0=100\,nm$ and $w=100\,nm$. Similar to 
the experimental data shown in Fig.~\ref{fig:iv-exp}, the model predicts very high rectification degrees with very 
small ion currents for reverse bias.

\subsection{\label{ss:pd}Position of depletion zone, dynamics of nanofluidic diode formation}

\begin{figure}[ht]
\includegraphics[width=0.48\textwidth]{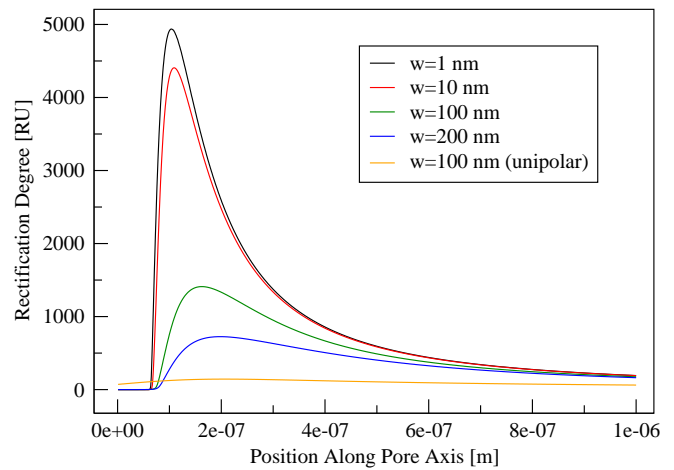}%
\caption{\label{fig:rectification}Rectification degree of a nanofluidic diode as a function of ``$r_0$'' parameter 
for various thicknesses of the transition zone ``$w$'' (Fig.\ \ref{fig:surf-charge}), at $2\,V$.}
\end{figure}

The process of preparation of nanofluidic diode is based on a chemical reaction of the reagent with the carboxyls 
groups on the pore walls. This causes change of the position $r_0$ and increase of the rectification degree in time 
when the chemical reaction progresses \cite{Vlassiouk:2007}.
\begin{figure}[ht]
\includegraphics[width=0.48\textwidth]{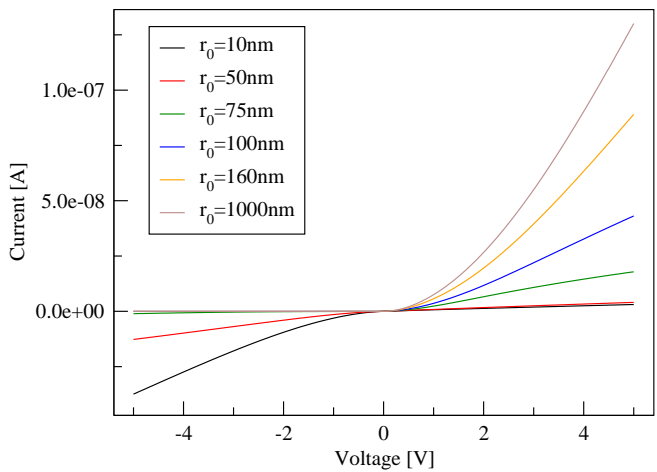}%
\caption{\label{fig:iv-curves}Current-voltage curves for a nanofluidic diode (Fig.\ \ref{fig:pore-geometry}) for 
various values of the $r_0$ parameter when $w=100\,nm$. Note the change of the direction of ion current 
rectification.}
\end{figure}
We decided to investigate in more details how sensitive the rectification degree is on the position of the transition 
zone in the pore. We performed therefore numerically the experiment of chemical modification of the pore walls. 
However, the reader must be aware that our surface charge distribution is not necessarily the distribution which 
corresponds to the real case. In the real case the shape of $\sigma(r)$ changes in time \cite{Vlassiouk:2007}, i.e., 
it changes with $r_0$, whereas in our case, the time evolution is simply described by a translation of $\sigma(r)$ 
along the radial coordinate. We think however that the main features of the real charge distribution are present in 
our model system and this gives us the possibility to understand and get insight into the nanofluidic diode behavior 
when the surface charge gets modified. A given value of the transition zone width $w$ was assumed and the position 
$r_0$ was moved between the pore tip and its base. In other words, we start from the situation where the whole 
surface of the pore is negatively charged and finish in a state where the whole surface is positively charged. In the 
initial and final states we have a homogeneously charged conical geometry that leads to ion current rectification via 
rocking ratchet mechanism described earlier in \cite{Siwy:2002fs} and \cite{Siwy:2004nr}. The rectification degree 
was defined in Eq.~\eqref{rectif-deg} as the ratio of currents recorded for positive voltages divided by currents 
recorded for negative voltages. These homogeneously charged conical nanopores are characterized by rectification 
degrees less than $10$. As we move the position $r_0$ inside the pore the surface charge in the tip changes acquiring 
more positive charges. As shown in Fig.~\ref{fig:iv-curves}, it is remarkable that a small amount of positive charge 
in the tip of the pore reverses the direction of rectification and changes the devices from monopolar to bipolar. The 
rectification degree versus $r_0$ is shown in Fig.~\ref{fig:rectification} for several values of the transition zone 
width.
The rectification degree increases steeply as we move $r_0$ inside the pore along its axis, reaches a maximum, and 
decreases to a value lower than $10$, when $r_0$ is located at the base of the pore. A very strong dependence of the 
rectification degree on $r_0$, when $r_0$ is located at the narrowest part of the pore, suggests that transport 
properties of the system are determined by the physical and chemical properties of the tip of the nanopore, in this 
case the surface charge distribution. It is also important to mention that we have predicted higher currents for 
devices where $r_0$ is located further from the tip (e.g. at $r_0=1\,\mu m$). However, these devices show lower 
rectification degrees due to larger leakage currents (off state currents). Depending on the specifics of the various 
applications nanofluidic diodes can therefore provide higher (lower) currents and lower (higher) rectification 
degrees. We would like to emphasize that all these various properties can be obtained by just changing the position 
of the zero surface charge point.

As a comparison, we have also included rectification data for a monopolar diode designed such that the tip of the 
pore was positively charged ($0.16\,C/m^2$) with the rest of the pore neutral ($0\,C/m^2$). The surface charge 
distribution is similar to the one shown in Fig.~\ref{fig:surf-charge}, the only difference is that $r_0$ represents 
in this case the position where the surface charge is half its maximum value. The parameter $w$ was set to $100\,nm$. 
One can see that overall the rectification degrees for the monopolar diode are much smaller than in case of the 
bipolar diode when both devices are studied in the same $KCl$ concentration.

Our results show that a huge increase of ion current rectification degrees is possible when the surface of a pore is 
patterned such that a part of a pore with negative surface charge is brought into contact with a part of the pore 
with positive surface charge. Our analysis also points to the importance of future experimental efforts aimed at a 
better control over the width of the transition zone as well as the position of the center of a depletion zone.

\subsection{\label{ss:sr}Scaling of rectification degree with voltage}

We have also analyzed the behavior of the rectification degree with voltage.
\begin{figure}[ht]
\includegraphics[width=0.48\textwidth]{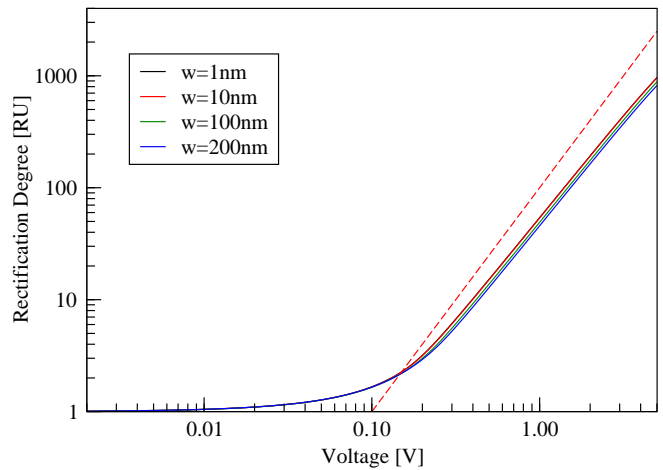}%
\caption{\label{fig:rectification-scaling-log}Log-log plot of the rectification degree of a nanofluidic diode 
(Fig.~\ref{fig:pore-geometry}) as a function of applied voltage at $r_0=1\mu m$ and for various values of the ``$w$'' 
parameter. The red dashed line is a power law with exponent equal to $2$.}
\end{figure}
In Fig.~\ref{fig:rectification-scaling-log} we show the rectification degree of a nanofluidic diode $f(V)$ as a 
function of voltage ranging between $-5\,V$ and $+5\,V$ for the case when the transition zone is located $1\,\mu m$ 
away from the tip, and for various values of $w$. We find that $f(V)$ has a power law dependence for voltages larger 
than $\approx\,0.3\,V$,
\begin{equation}
\label{v-scaling}
f(V) \propto V^\gamma\,,
\end{equation}
where the scaling exponent $\gamma\approx1.81$. For guidance, the power law with exponent equal to $2$ is shown in 
Fig.~\ref{fig:rectification-scaling-log}. The power law scaling of the rectification degree with voltage might 
suggest the existence of universal scaling. To our knowledge, this is the first time when this scaling has been 
observed in the variation of the rectification degree with the applied voltage for a nanofluidic diode based on a 
single conical nanopore. The applicability of the universal scaling and universal scaling exponents to nanofluidic 
diodes will be investigated in our future studies.

\subsection{\label{ss:dd}Dependence of depletion zone width on voltage}

\begin{figure}[ht]
\includegraphics[width=0.48\textwidth]{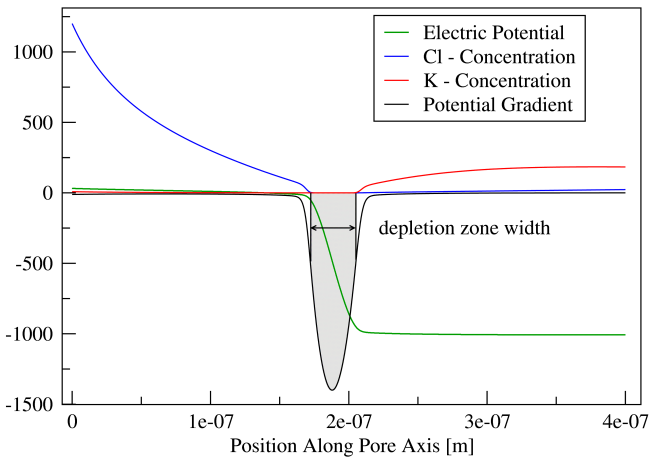}%
\caption{\label{fig:electric-field}Profiles of ionic concentrations, electric potential and its first derivative at 
the depletion zone created at the reverse bias of $-2\,V$. Values have been scaled (see text) to make the graphic 
visualization possible.}
\end{figure}

Another interesting question to consider is the dependence of the depletion zone width $w_d$ on the reverse bias 
voltages. The depletion zone is defined as the region where mobile charges are absent or at least their number is 
small compared to adjacent regions Fig.~\ref{fig:profiles}. This ion free region appears only in the reverse bias 
regime. The lack of ions in this zone produces a drop of the electric potential, or in other words the electric field 
in this region is strong enough to block the migration of mobile charges through the pore. To help visualize the 
above statements we have superimposed in Fig.~\ref{fig:electric-field} the concentration profiles, the electric 
potential, and its first derivative which is proportional to the electric field. Since the voltage drops over a very 
small distance the electric fields created are of the order of magnitude $10^8\,V/m$  .
As we have explained above, the electric field has a strong peak in the depletion zone and we have considered the 
width of this peak to provide the definition for the width of the depletion zone. In order to plot all these profiles 
in one figure, scaling of their values was necessary. The electric potential was scaled up $50$ times and its first 
derivative was scaled down $5000$ times such that all the graphs will provide sufficient details. The plots in 
Fig.~\ref{fig:electric-field} correspond to $w=100\,nm$, $r_0=250\,nm$, and for a reverse bias voltage of $-2\,V$.

\begin{figure}[ht]
\includegraphics[width=0.48\textwidth]{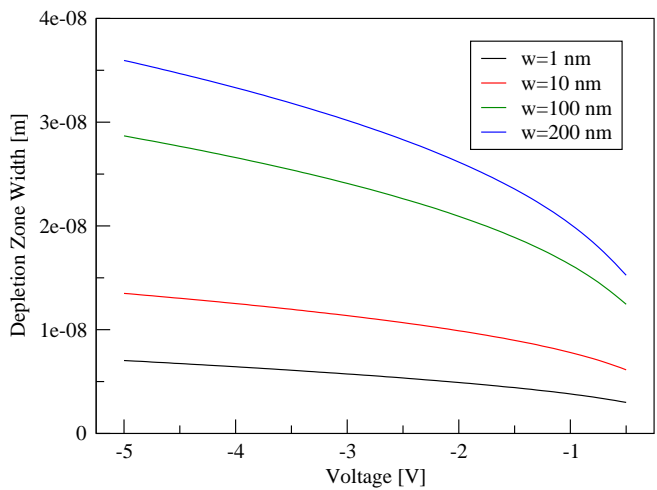}%
\caption{\label{fig:depletion-width}Dependence of depletion zone width $w_d$ on voltage for various values of the 
width of the transition zone $w$.}
\end{figure}

We would like to emphasize the difference between the width of the transition zone $w$ and the width of the depletion 
zone. Larger $w$ values lead to larger values of $w_d$, as shown in Fig.~\ref{fig:depletion-width}.
For $w$ equal to $1\,nm$ and $10\,nm$, the resulting $w_d$ for $5\,V$ exceeds the value of the transition zone width 
that assured formation of much smaller leakage currents, and consequently higher rectification degrees. For 
transition zone of width $100\,nm$ and $200\,nm$, the depletion zone is narrower than the values of $w$. Larger $w_d$ 
therefore does not necessarily imply higher rectification degrees, as explained in the previous section. Smaller 
values of $w$ and $w_d$ assure larger electric fields for the reverse bias, which also helps formation of a depletion 
zone of a larger resistance.


\section{\label{s:conclusions}Conclusions}


The modeling of a nanofluidic diode presented in this manuscript has been motivated and inspired by recent 
experimental realization of this device \cite{Vlassiouk:2007}. In this article we have shown that rectification 
degrees of $10^3$ can be achieved with this device upon more precise surface charge control. We have shown that the 
width of the transition zone and the position of the transition point are parameters which influence very strongly 
the rectification properties of nanofluidic diodes. These parameters in turn are controlled by chemical modification 
of pore walls. Our analysis also shows that preparation of a nanofluidic diode based on a conical geometry offers 
much freedom in tuning rectification properties of the device. Future experimental efforts will be focused on 
developing surface chemistries so that various rectification degrees can be obtained. Our modeling also provide 
directions for the future design of the nanofluidic diode in order to prepare versatile devices which meet the 
requirements of various applications.

It would be very interesting to see how the rectification behaves when other parameters of the problem are changed, 
for example shape of the nanopore, and magnitude of surface charge. Another interesting aspect of the bipolar diode 
is the power law dependence of rectification degrees on voltage, which might suggest existence of universal scaling 
and critical exponents in the system. Our future studies will be directed towards identification of these critical 
exponents and relating them to formation of the depletion zone.


\begin{acknowledgments}
We are grateful to Dr. Ivan Vlassiouk for stimulating discussions, and Dr. Magdalena Constantin for careful reading 
of our manuscript. We thank the Alfred P. Sloan Foundation and the
Institute for Complex Adaptive Matter for financial support.
\end{acknowledgments}


\bibliographystyle{apsrev}

\begin{thebibliography}{34}
\expandafter\ifx\csname natexlab\endcsname\relax\def\natexlab#1{#1}\fi
\expandafter\ifx\csname bibnamefont\endcsname\relax
  \def\bibnamefont#1{#1}\fi
\expandafter\ifx\csname bibfnamefont\endcsname\relax
  \def\bibfnamefont#1{#1}\fi
\expandafter\ifx\csname citenamefont\endcsname\relax
  \def\citenamefont#1{#1}\fi
\expandafter\ifx\csname url\endcsname\relax
  \def\url#1{\texttt{#1}}\fi
\expandafter\ifx\csname urlprefix\endcsname\relax\def\urlprefix{URL }\fi
\providecommand{\bibinfo}[2]{#2}
\providecommand{\eprint}[2][]{\url{#2}}

\bibitem[{\citenamefont{Lev et~al.}(1993)\citenamefont{Lev, Korchev,
  Rostovtseva, Bashford, Edmonds, and Pasternak}}]{Lev:1993rs}
\bibinfo{author}{\bibfnamefont{A.~A.} \bibnamefont{Lev}},
  \bibinfo{author}{\bibfnamefont{Y.~E.} \bibnamefont{Korchev}},
  \bibinfo{author}{\bibfnamefont{T.~K.} \bibnamefont{Rostovtseva}},
  \bibinfo{author}{\bibfnamefont{C.~L.} \bibnamefont{Bashford}},
  \bibinfo{author}{\bibfnamefont{D.~T.} \bibnamefont{Edmonds}},
  \bibnamefont{and} \bibinfo{author}{\bibfnamefont{C.~A.}
  \bibnamefont{Pasternak}}, \bibinfo{journal}{Proceedings: Biological Sciences}
  \textbf{\bibinfo{volume}{{\bf 252}}}, \bibinfo{pages}{187}
  (\bibinfo{year}{1993}).

\bibitem[{\citenamefont{Siwy et~al.}(2003)\citenamefont{Siwy, Apel, Baur,
  Dobrev, Korchev, Neumann, Spohr, Trautmann, and Voss}}]{Siwy:2003ps}
\bibinfo{author}{\bibfnamefont{Z.}~\bibnamefont{Siwy}},
  \bibinfo{author}{\bibfnamefont{P.}~\bibnamefont{Apel}},
  \bibinfo{author}{\bibfnamefont{D.}~\bibnamefont{Baur}},
  \bibinfo{author}{\bibfnamefont{D.~D.} \bibnamefont{Dobrev}},
  \bibinfo{author}{\bibfnamefont{Y.~E.} \bibnamefont{Korchev}},
  \bibinfo{author}{\bibfnamefont{R.}~\bibnamefont{Neumann}},
  \bibinfo{author}{\bibfnamefont{R.}~\bibnamefont{Spohr}},
  \bibinfo{author}{\bibfnamefont{C.}~\bibnamefont{Trautmann}},
  \bibnamefont{and} \bibinfo{author}{\bibfnamefont{K.-O.} \bibnamefont{Voss}},
  \bibinfo{journal}{Surf. Sci.} \textbf{\bibinfo{volume}{{\bf 532-535}}},
  \bibinfo{pages}{1061} (\bibinfo{year}{2003}).

\bibitem[{\citenamefont{Chen et~al.}(2004)\citenamefont{Chen, Mitsui, Farmer,
  Golovchenko, Gordon, and Branton}}]{Chen:2004al}
\bibinfo{author}{\bibfnamefont{P.}~\bibnamefont{Chen}},
  \bibinfo{author}{\bibfnamefont{T.}~\bibnamefont{Mitsui}},
  \bibinfo{author}{\bibfnamefont{D.~B.} \bibnamefont{Farmer}},
  \bibinfo{author}{\bibfnamefont{J.}~\bibnamefont{Golovchenko}},
  \bibinfo{author}{\bibfnamefont{R.~G.} \bibnamefont{Gordon}},
  \bibnamefont{and} \bibinfo{author}{\bibfnamefont{D.}~\bibnamefont{Branton}},
  \bibinfo{journal}{Nano Letters} \textbf{\bibinfo{volume}{{\bf 4}}},
  \bibinfo{pages}{1333} (\bibinfo{year}{2004}).

\bibitem[{\citenamefont{Li et~al.}(2001)\citenamefont{Li, Stein, McMullan,
  Branton, Aziz, and Golovchenko}}]{Li:2001ib}
\bibinfo{author}{\bibfnamefont{J.}~\bibnamefont{Li}},
  \bibinfo{author}{\bibfnamefont{D.}~\bibnamefont{Stein}},
  \bibinfo{author}{\bibfnamefont{C.}~\bibnamefont{McMullan}},
  \bibinfo{author}{\bibfnamefont{D.}~\bibnamefont{Branton}},
  \bibinfo{author}{\bibfnamefont{M.~J.} \bibnamefont{Aziz}}, \bibnamefont{and}
  \bibinfo{author}{\bibfnamefont{J.~A.} \bibnamefont{Golovchenko}},
  \bibinfo{journal}{Nature} \textbf{\bibinfo{volume}{{\bf 412}}},
  \bibinfo{pages}{166} (\bibinfo{year}{2001}).

\bibitem[{\citenamefont{Dekker}(2007)}]{Dekker:2007ss}
\bibinfo{author}{\bibfnamefont{C.}~\bibnamefont{Dekker}},
  \bibinfo{journal}{Nature Nanotech.} \textbf{\bibinfo{volume}{{\bf 2}}},
  \bibinfo{pages}{209} (\bibinfo{year}{2007}).

\bibitem[{\citenamefont{Harrell et~al.}(2006)\citenamefont{Harrell, Choi,
  Horne, Baker, Siwy, and Martin}}]{Harrell:2006rp}
\bibinfo{author}{\bibfnamefont{C.~C.} \bibnamefont{Harrell}},
  \bibinfo{author}{\bibfnamefont{Y.}~\bibnamefont{Choi}},
  \bibinfo{author}{\bibfnamefont{L.~P.} \bibnamefont{Horne}},
  \bibinfo{author}{\bibfnamefont{L.~A.} \bibnamefont{Baker}},
  \bibinfo{author}{\bibfnamefont{Z.~S.} \bibnamefont{Siwy}}, \bibnamefont{and}
  \bibinfo{author}{\bibfnamefont{C.~R.} \bibnamefont{Martin}},
  \bibinfo{journal}{Langmuir} \textbf{\bibinfo{volume}{{\bf 22}}},
  \bibinfo{pages}{10837} (\bibinfo{year}{2006}).

\bibitem[{\citenamefont{Mara et~al.}(2004)\citenamefont{Mara, Siwy, Trautmann,
  Wan, and Kamme}}]{Mara:2004aa}
\bibinfo{author}{\bibfnamefont{A.}~\bibnamefont{Mara}},
  \bibinfo{author}{\bibfnamefont{Z.}~\bibnamefont{Siwy}},
  \bibinfo{author}{\bibfnamefont{C.}~\bibnamefont{Trautmann}},
  \bibinfo{author}{\bibfnamefont{J.}~\bibnamefont{Wan}}, \bibnamefont{and}
  \bibinfo{author}{\bibfnamefont{F.}~\bibnamefont{Kamme}},
  \bibinfo{journal}{Nano Letters} \textbf{\bibinfo{volume}{{\bf 4}}},
  \bibinfo{pages}{497} (\bibinfo{year}{2004}).

\bibitem[{\citenamefont{Gijs}(2007)}]{Gijs:2007}
\bibinfo{author}{\bibfnamefont{M.~A.~M.} \bibnamefont{Gijs}},
  \bibinfo{journal}{Nature Nanotech.} \textbf{\bibinfo{volume}{{\bf 2}}},
  \bibinfo{pages}{268} (\bibinfo{year}{2007}).

\bibitem[{\citenamefont{Vlassiouk and Siwy}(2007)}]{Vlassiouk:2007}
\bibinfo{author}{\bibfnamefont{I.}~\bibnamefont{Vlassiouk}} \bibnamefont{and}
  \bibinfo{author}{\bibfnamefont{Z.~S.} \bibnamefont{Siwy}},
  \bibinfo{journal}{Nano Letters} \textbf{\bibinfo{volume}{{\bf 7}}},
  \bibinfo{pages}{552} (\bibinfo{year}{2007}).

\bibitem[{\citenamefont{Karnik et~al.}(2007)\citenamefont{Karnik, Duan,
  Castelino, Daiguji, and Majumdar}}]{Karnik:2007ri}
\bibinfo{author}{\bibfnamefont{R.}~\bibnamefont{Karnik}},
  \bibinfo{author}{\bibfnamefont{C.}~\bibnamefont{Duan}},
  \bibinfo{author}{\bibfnamefont{K.}~\bibnamefont{Castelino}},
  \bibinfo{author}{\bibfnamefont{H.}~\bibnamefont{Daiguji}}, \bibnamefont{and}
  \bibinfo{author}{\bibfnamefont{A.}~\bibnamefont{Majumdar}},
  \bibinfo{journal}{Nano Letters} \textbf{\bibinfo{volume}{{\bf 7}}},
  \bibinfo{pages}{547} (\bibinfo{year}{2007}).

\bibitem[{\citenamefont{Schuss et~al.}(2001)\citenamefont{Schuss, Nadler, and
  Eisenberg}}]{Schuss:2001dp}
\bibinfo{author}{\bibfnamefont{Z.}~\bibnamefont{Schuss}},
  \bibinfo{author}{\bibfnamefont{B.}~\bibnamefont{Nadler}}, \bibnamefont{and}
  \bibinfo{author}{\bibfnamefont{R.~S.} \bibnamefont{Eisenberg}},
  \bibinfo{journal}{Phys. Rev. E} \textbf{\bibinfo{volume}{{\bf 64}}},
  \bibinfo{pages}{036116} (\bibinfo{year}{2001}).

\bibitem[{\citenamefont{Nadler et~al.}(2004)\citenamefont{Nadler, Schuss,
  Singer, and Eisenberg}}]{Nadler:2004id}
\bibinfo{author}{\bibfnamefont{B.}~\bibnamefont{Nadler}},
  \bibinfo{author}{\bibfnamefont{Z.}~\bibnamefont{Schuss}},
  \bibinfo{author}{\bibfnamefont{A.}~\bibnamefont{Singer}}, \bibnamefont{and}
  \bibinfo{author}{\bibfnamefont{R.~S.} \bibnamefont{Eisenberg}},
  \bibinfo{journal}{J. Phys.: Condens. Matter} \textbf{\bibinfo{volume}{{\bf
  16}}}, \bibinfo{pages}{S2153} (\bibinfo{year}{2004}).

\bibitem[{\citenamefont{Chen et~al.}(1997)\citenamefont{Chen, Lear, and
  Eisenberg}}]{Chen:1997pt}
\bibinfo{author}{\bibfnamefont{D.}~\bibnamefont{Chen}},
  \bibinfo{author}{\bibfnamefont{J.}~\bibnamefont{Lear}}, \bibnamefont{and}
  \bibinfo{author}{\bibfnamefont{R.~S.} \bibnamefont{Eisenberg}},
  \bibinfo{journal}{Biophys. J.} \textbf{\bibinfo{volume}{{\bf 72}}},
  \bibinfo{pages}{97} (\bibinfo{year}{1997}).

\bibitem[{\citenamefont{Liu et~al.}(2007)\citenamefont{Liu, Wang, Guo, Ji, Xue,
  and Ouyang}}]{Liu:2007ap}
\bibinfo{author}{\bibfnamefont{Q.}~\bibnamefont{Liu}},
  \bibinfo{author}{\bibfnamefont{Y.}~\bibnamefont{Wang}},
  \bibinfo{author}{\bibfnamefont{W.}~\bibnamefont{Guo}},
  \bibinfo{author}{\bibfnamefont{H.}~\bibnamefont{Ji}},
  \bibinfo{author}{\bibfnamefont{J.}~\bibnamefont{Xue}}, \bibnamefont{and}
  \bibinfo{author}{\bibfnamefont{Q.}~\bibnamefont{Ouyang}},
  \bibinfo{journal}{Phys. Rev. E} \textbf{\bibinfo{volume}{{\bf 75}}},
  \bibinfo{pages}{051201} (\bibinfo{year}{2007}).

\bibitem[{\citenamefont{Cervera et~al.}(2005)\citenamefont{Cervera, Schiedt,
  and Ram\'{\i}rez}}]{Cervera:2005}
\bibinfo{author}{\bibfnamefont{J.}~\bibnamefont{Cervera}},
  \bibinfo{author}{\bibfnamefont{B.}~\bibnamefont{Schiedt}}, \bibnamefont{and}
  \bibinfo{author}{\bibfnamefont{P.}~\bibnamefont{Ram\'{\i}rez}},
  \bibinfo{journal}{Europhys. Lett.} \textbf{\bibinfo{volume}{{\bf 71}}},
  \bibinfo{pages}{35} (\bibinfo{year}{2005}).

\bibitem[{\citenamefont{Cervera et~al.}(2006)\citenamefont{Cervera, Schiedt,
  Neumann, Maf\'e, and Ram\'{\i}rez}}]{Cervera:2006}
\bibinfo{author}{\bibfnamefont{J.}~\bibnamefont{Cervera}},
  \bibinfo{author}{\bibfnamefont{B.}~\bibnamefont{Schiedt}},
  \bibinfo{author}{\bibfnamefont{R.}~\bibnamefont{Neumann}},
  \bibinfo{author}{\bibfnamefont{S.}~\bibnamefont{Maf\'e}}, \bibnamefont{and}
  \bibinfo{author}{\bibfnamefont{P.}~\bibnamefont{Ram\'{\i}rez}},
  \bibinfo{journal}{J. Chem. Phys.} \textbf{\bibinfo{volume}{{\bf 124}}},
  \bibinfo{pages}{104706} (\bibinfo{year}{2006}).

\bibitem[{\citenamefont{Daiguji et~al.}(2005)\citenamefont{Daiguji, Oka, and
  Shirono}}]{Daiguji:2005nd}
\bibinfo{author}{\bibfnamefont{H.}~\bibnamefont{Daiguji}},
  \bibinfo{author}{\bibfnamefont{Y.}~\bibnamefont{Oka}}, \bibnamefont{and}
  \bibinfo{author}{\bibfnamefont{K.}~\bibnamefont{Shirono}},
  \bibinfo{journal}{Nano Letters} \textbf{\bibinfo{volume}{{\bf 5}}},
  \bibinfo{pages}{2274} (\bibinfo{year}{2005}).

\bibitem[{\citenamefont{Daiguji et~al.}(2004)\citenamefont{Daiguji, Yang, and
  Majumdar}}]{Daiguji:2004it}
\bibinfo{author}{\bibfnamefont{H.}~\bibnamefont{Daiguji}},
  \bibinfo{author}{\bibfnamefont{P.}~\bibnamefont{Yang}}, \bibnamefont{and}
  \bibinfo{author}{\bibfnamefont{A.}~\bibnamefont{Majumdar}},
  \bibinfo{journal}{Nano Letters} \textbf{\bibinfo{volume}{{\bf 4}}},
  \bibinfo{pages}{137} (\bibinfo{year}{2004}).

\bibitem[{\citenamefont{Nonner and Eisenberg}(1998)}]{Nonner:1998}
\bibinfo{author}{\bibfnamefont{W.}~\bibnamefont{Nonner}} \bibnamefont{and}
  \bibinfo{author}{\bibfnamefont{B.}~\bibnamefont{Eisenberg}},
  \bibinfo{journal}{Biophys. J.} \textbf{\bibinfo{volume}{{\bf 75}}},
  \bibinfo{pages}{1287} (\bibinfo{year}{1998}).

\bibitem[{\citenamefont{Corry et~al.}(2000)\citenamefont{Corry, Kuyucak, and
  Chung}}]{Corry:2000}
\bibinfo{author}{\bibfnamefont{B.}~\bibnamefont{Corry}},
  \bibinfo{author}{\bibfnamefont{S.}~\bibnamefont{Kuyucak}}, \bibnamefont{and}
  \bibinfo{author}{\bibfnamefont{S.~H.} \bibnamefont{Chung}},
  \bibinfo{journal}{Biophys. J.} \textbf{\bibinfo{volume}{{\bf 78}}},
  \bibinfo{pages}{2364} (\bibinfo{year}{2000}).

\bibitem[{\citenamefont{Apel et~al.}(2001)\citenamefont{Apel, Korchev, Siwy,
  Spohr, and Yoshida}}]{Apel:2001dl}
\bibinfo{author}{\bibfnamefont{P.~Y.} \bibnamefont{Apel}},
  \bibinfo{author}{\bibfnamefont{Y.~E.} \bibnamefont{Korchev}},
  \bibinfo{author}{\bibfnamefont{Z.}~\bibnamefont{Siwy}},
  \bibinfo{author}{\bibfnamefont{R.}~\bibnamefont{Spohr}}, \bibnamefont{and}
  \bibinfo{author}{\bibfnamefont{M.}~\bibnamefont{Yoshida}},
  \bibinfo{journal}{Nucl. Instrum. and Met. in Phys. Res.}
  \textbf{\bibinfo{volume}{{\bf B 184}}}, \bibinfo{pages}{337}
  (\bibinfo{year}{2001}).

\bibitem[{\citenamefont{Siwy}(2006)}]{Siwy:2006ic}
\bibinfo{author}{\bibfnamefont{Z.}~\bibnamefont{Siwy}}, \bibinfo{journal}{Adv.
  Funct. Mater.} \textbf{\bibinfo{volume}{{\bf 16}}}, \bibinfo{pages}{735}
  (\bibinfo{year}{2006}).

\bibitem[{\citenamefont{Spohr}(1983)}]{Spohrpatent}
\bibinfo{author}{\bibfnamefont{R.}~\bibnamefont{Spohr}},
  \emph{\bibinfo{title}{{\em `` Method for Producing Nuclear Traces or
  Microholes Originating from Nuclear Traces of an Individual Ion''}}}
  (\bibinfo{year}{1983}), \bibinfo{note}{{\em ``German Patent DE 2951376 C2;
  United States Patent 4369370''}}.

\bibitem[{\citenamefont{Wolf-Reber}(2002)}]{Wolf}
\bibinfo{author}{\bibfnamefont{A.}~\bibnamefont{Wolf-Reber}}, Ph.D. thesis
  (\bibinfo{year}{2002}).

\bibitem[{\citenamefont{Grabarek and Gergely}(1990)}]{Grabarek:1990zl}
\bibinfo{author}{\bibfnamefont{Z.}~\bibnamefont{Grabarek}} \bibnamefont{and}
  \bibinfo{author}{\bibfnamefont{J.}~\bibnamefont{Gergely}},
  \bibinfo{journal}{Anal. Biochem.} \textbf{\bibinfo{volume}{{\bf 185}}},
  \bibinfo{pages}{131} (\bibinfo{year}{1990}).

\bibitem[{\citenamefont{Coster}(1965)}]{Coster:1965qa}
\bibinfo{author}{\bibfnamefont{H.~G.~L.} \bibnamefont{Coster}},
  \bibinfo{journal}{Biophys. J.} \textbf{\bibinfo{volume}{{\bf 5}}},
  \bibinfo{pages}{669} (\bibinfo{year}{1965}).

\bibitem[{\citenamefont{Bassignana and Reiss}(1983)}]{Bassignana:1983it}
\bibinfo{author}{\bibfnamefont{I.~C.} \bibnamefont{Bassignana}}
  \bibnamefont{and} \bibinfo{author}{\bibfnamefont{H.}~\bibnamefont{Reiss}},
  \bibinfo{journal}{J. Membr. Sci.} \textbf{\bibinfo{volume}{{\bf 15}}},
  \bibinfo{pages}{27} (\bibinfo{year}{1983}).

\bibitem[{\citenamefont{Maf\'e and Ram\'{\i}rez}(1997)}]{Mafe:1997ec}
\bibinfo{author}{\bibfnamefont{S.}~\bibnamefont{Maf\'e}} \bibnamefont{and}
  \bibinfo{author}{\bibfnamefont{P.}~\bibnamefont{Ram\'{\i}rez}},
  \bibinfo{journal}{Acta Polym.} \textbf{\bibinfo{volume}{{\bf 48}}},
  \bibinfo{pages}{234} (\bibinfo{year}{1997}).

\bibitem[{\citenamefont{Siwy et~al.}(2006{\natexlab{a}})\citenamefont{Siwy,
  Powell, Kalman, Astumian, and Eisenberg}}]{Siwy:2006ni}
\bibinfo{author}{\bibfnamefont{Z.}~\bibnamefont{Siwy}},
  \bibinfo{author}{\bibfnamefont{M.~R.} \bibnamefont{Powell}},
  \bibinfo{author}{\bibfnamefont{E.}~\bibnamefont{Kalman}},
  \bibinfo{author}{\bibfnamefont{R.~D.} \bibnamefont{Astumian}},
  \bibnamefont{and} \bibinfo{author}{\bibfnamefont{R.~S.}
  \bibnamefont{Eisenberg}}, \bibinfo{journal}{Nano Letters}
  \textbf{\bibinfo{volume}{{\bf 6}}}, \bibinfo{pages}{473}
  (\bibinfo{year}{2006}{\natexlab{a}}).

\bibitem[{\citenamefont{Siwy et~al.}(2006{\natexlab{b}})\citenamefont{Siwy,
  Powell, Petrov, Kalman, Trautmann, and Eisenberg}}]{Siwy:2006ci}
\bibinfo{author}{\bibfnamefont{Z.}~\bibnamefont{Siwy}},
  \bibinfo{author}{\bibfnamefont{M.~R.} \bibnamefont{Powell}},
  \bibinfo{author}{\bibfnamefont{A.}~\bibnamefont{Petrov}},
  \bibinfo{author}{\bibfnamefont{E.}~\bibnamefont{Kalman}},
  \bibinfo{author}{\bibfnamefont{C.}~\bibnamefont{Trautmann}},
  \bibnamefont{and} \bibinfo{author}{\bibfnamefont{R.~S.}
  \bibnamefont{Eisenberg}}, \bibinfo{journal}{Nano Letters}
  \textbf{\bibinfo{volume}{{\bf 6}}}, \bibinfo{pages}{1729}
  (\bibinfo{year}{2006}{\natexlab{b}}).

\bibitem[{\citenamefont{Gillespie and Eisenberg}(2001)}]{Gillespie:2001}
\bibinfo{author}{\bibfnamefont{D.}~\bibnamefont{Gillespie}} \bibnamefont{and}
  \bibinfo{author}{\bibfnamefont{R.~S.} \bibnamefont{Eisenberg}},
  \bibinfo{journal}{Phys. Rev. E} \textbf{\bibinfo{volume}{{\bf 63}}},
  \bibinfo{pages}{061902} (\bibinfo{year}{2001}).

\bibitem[{\citenamefont{Shampine et~al.}(2006)\citenamefont{Shampine, Muir, and
  Xu}}]{Shampine:2006}
\bibinfo{author}{\bibfnamefont{L.}~\bibnamefont{Shampine}},
  \bibinfo{author}{\bibfnamefont{P.}~\bibnamefont{Muir}}, \bibnamefont{and}
  \bibinfo{author}{\bibfnamefont{H.}~\bibnamefont{Xu}},
  \bibinfo{journal}{JNAIAM} \textbf{\bibinfo{volume}{{\bf 1}}},
  \bibinfo{pages}{201} (\bibinfo{year}{2006}).

\bibitem[{\citenamefont{Siwy and Fuli\ifmmode~\acute{n}\else
  \'{n}\fi{}ski}(2002)}]{Siwy:2002fs}
\bibinfo{author}{\bibfnamefont{Z.}~\bibnamefont{Siwy}} \bibnamefont{and}
  \bibinfo{author}{\bibfnamefont{A.}~\bibnamefont{Fuli\ifmmode~\acute{n}\else
  \'{n}\fi{}ski}}, \bibinfo{journal}{Phys. Rev. Lett.}
  \textbf{\bibinfo{volume}{{\bf 89}}}, \bibinfo{pages}{198103}
  (\bibinfo{year}{2002}).

\bibitem[{\citenamefont{Siwy and Fuli\ifmmode~\acute{n}\else
  \'{n}\fi{}ski}(2004)}]{Siwy:2004nr}
\bibinfo{author}{\bibfnamefont{Z.}~\bibnamefont{Siwy}} \bibnamefont{and}
  \bibinfo{author}{\bibfnamefont{A.}~\bibnamefont{Fuli\ifmmode~\acute{n}\else
  \'{n}\fi{}ski}}, \bibinfo{journal}{Am. J. Phys.}
  \textbf{\bibinfo{volume}{{\bf 72}}}, \bibinfo{pages}{567}
  (\bibinfo{year}{2004}).

\end{thebibliography}



\end{document}